# MST3 Encryption improvement with three-parameter group of Hermitian function field

Gennady Khalimov[1[0000-0002-2054-9186]], Yevgen Kotukh [2[0000-0003-4997-620X]]

[1]Kharkiv National University of Radioelectronics, Kharkiv, 61166, Ukraine hennadii.khalimov@nure.ua
[2]Yevhenii Bereznyak Military Academy, Kyiv, Ukraine yevgenkotukh@gmail.com

**Abstract.** This scholarly work presents an advanced cryptographic framework utilizing automorphism groups as the foundational structure for encryption scheme implementation. The proposed methodology employs a three-parameter group construction, distinguished by its application of logarithmic signatures positioned outside the group's center—a significant departure from conventional approaches. A key innovation in this implementation is utilizing the Hermitian function field as the underlying mathematical framework. This particular function field provides enhanced structural properties that strengthen the cryptographic protocol when integrated with the three-parameter group architecture. The encryption mechanism features phased key de-encapsulation from ciphertext, representing a substantial advantage over alternative implementations. This sequential extraction process introduces additional computational complexity for potential adversaries while maintaining efficient legitimate decryption. A notable characteristic of this cryptosystem is the direct correlation between the underlying group's mathematical strength and both the attack complexity and message size parameters. This relationship enables precise security-efficiency calibration based on specific implementation requirements and threat models. The application of automorphism groups with logarithmic signatures positioned outside the center represents a significant advancement in non-traditional cryptographic designs, particularly relevant in the context of post-quantum cryptographic resilience.

# INTRODUCTION

The imminent development of large-scale quantum computing technology poses a substantial threat to contemporary public key cryptosystems. Specifically, cryptographic protocols predicated on integer factorization complexity or discrete logarithm problems, including widely deployed systems such as RSA and ECC, would be rendered vulnerable. Over approximately two decades, researchers have proposed several approaches utilizing non-commutative groups to construct quantum-resistant cryptosystems [1-4].

The intractable word problem represents a particularly promising research direction for cryptographic system development. Initially formulated by Wagner and Magyarik [5], this approach functions within the domain of permutation group applications. Magliveras [6] introduced logarithmic signatures — specialized factorization structures applicable to finite groups, establishing the foundation for subsequent cryptographic protocols. Refinements to the original methodology were developed [7-9], culminating in the MST3 implementation [9] based on Suzuki group properties.

In 2008, Magliveras et al. [10] identified limitations associated with transitive logarithmic signature utilization in MST3 cryptosystems. Subsequently, Svaba et al. [11] proposed an enhanced variant, designated eMST3, featuring improved security through the integration of secret homomorphic covers. Further advancement occurred in 2018 when T. van Trung [12] developed an MST3 approach utilizing strong aperiodic logarithmic signatures specifically for Abelian p-groups.

Y. Cong et al. [13] conducted comprehensive analysis of MST3 implementations, noting that the absence of published research demonstrating quantum vulnerability positions these algorithms as viable candidates for post-quantum cryptographic applications.

The original approach within MST3 cryptosystem construction is based on the Suzuki group. There are several approaches for further improvements of MST3 were considered [14–17]. One of the valuable ideas is to increase the encryption efficiency by optimizing calculation overheads. It was done with reduction of large size of the keys space. Authors shown that the approach to apply for the LS computing outside of the group center. And it was done over finite fields of small dimension using groups with large order. Suzuki groups are isomorphic to the projective linear group $PGL(3, F_q)$, where $q = 2q_0^2$, $q_0 = 2^n$ and has order $q^2$. Basically, cryptosystem security is defined by group order. In [16], authors proposed three-parameter group of the automorphism for the first time. It applies to construct MST3 cryptosystem with improved security. $H(P_\infty)$ of the $Herm|F_{q^2}$ has a greater $ordH(P_\infty) = q^3(q^2 - 1)$ and greater than the order of corresponding Suzuki group being considered at original papers. Our paper presents a practical implementation of this new approach.

# THREE-PARAMETER AUTOMORPHISM GROUP OF THE HERMITIAN FUNCTION FIELD

The $Herm|F_{q^2}$ in [14]. We use $Aut(Herm)$ of the $Herm|F_{q^2}$ that can be presented as follows $H := Aut(Herm) = \{\psi : Herm \mapsto Herm | \psi$ of $Herm|F_{q^2}\}$

And it's extremely large [14]. The properties were discussed in [14]. This group has the order equal to $ordA = q^3(q^2-1)(q^3+1)$. The decomposition group $H(P_\infty)$ has got of all $Aut(Herm)$ of $Herm|F_{q^2}$ with the following properties

$$\begin{cases} \psi(y) = \alpha y + \beta \\ \psi(z) = \alpha^{q+1} z + \alpha \beta^q y + \gamma, \end{cases}$$

where $\alpha \in F_{q^2}^* := F_{q^2} \setminus \{0\}$, $\beta \in F_{q^2}$ and $\gamma^q + \gamma = \beta^{q+1}$.

It has order $ordH(P_\infty) = q^3(q^2-1)$.

Here we have structure for the group as follows

$$[\alpha_1, \beta_1, \gamma_1] \cdot [\alpha_2, \beta_2, \gamma_2] = [\alpha_1 \alpha_2, \alpha_2 \beta_1 + \beta_2, \alpha_2^{q+1} \gamma_1 + \alpha_2 \beta_2^q \beta_1 + \gamma_2].$$

Thus, we have identity $[1,0,0]$ and the inverse of $[\alpha, \beta, \gamma]$ is

$$[\alpha, \beta, \gamma]^{-1} = [\alpha^{-1}, -\alpha^{-1}\beta, \alpha^{-(q+1)}\gamma^q].$$

$H(P_\infty)$ can be represented in simpler way by due to the off characteristic:

$$H(P_\infty)\left\{\left[\alpha, \beta, \frac{\beta^{q+1}}{2} + \gamma\right] \middle| \alpha \in F_{q^2}^*, \beta \in F_{q^2}, \gamma^q + \gamma = 0\right\}.$$

The unique $p$-Sylow subgroup of $H(P_\infty)$ can be designated by $H_1(P_\infty)$ within the following representation $H_1(P_\infty) = \{\psi \in H(P_\infty) | \psi(y) = y + \beta$ for some $b \in F_{q^2}\}$.

In such a case we have an order equal to $q^3$ for the unique $p$-Sylow subgroup of

$$\begin{cases} \psi(y) = y + \beta \\ \psi(z) = z + \beta^q y + \gamma, \end{cases}$$

where, $\beta \in F_{q^2}$, $\gamma^q + \gamma = \beta^{q+1}$ and its order equal to $q^3$ as we mentioned above. The structure for the group can be achieved by subgroup of $PGL(3,k)$ presentation:

$$H_1 := \left\{\begin{pmatrix} 1 & \beta & \gamma \\ 0 & 1 & \beta^q \\ 0 & 0 & 1 \end{pmatrix}, \gamma \in F_{q^2}, \gamma^q + \gamma = \beta^{q+1}\right\}$$

Group operation is defined as

$$[1,\beta_1,\gamma_1]\cdot[1,\beta_2,\gamma_2]=[1,\beta_1+\beta_2,\gamma_1+\beta_2^q\beta_1+\gamma_2]$$

and

$$[\beta_1,\gamma_1]\cdot[\beta_2,\gamma_2]=[\beta_1+\beta_2,\gamma_1+\beta_2^q\beta_1+\gamma_2].$$

The factor group $H(P_\infty)/H_1(P_\infty)$ is cyclic of order $q^2-1$. Moreover, it was generated by the $\varsigma \in H(P_\infty)$ with $\varsigma(y)=\alpha z,\ \varsigma(z)=\alpha^{q+1}z$.

Another automorphism $\zeta \in H$ is given by $\zeta(y)=y/z,\ \zeta(z)=1/z$.

The automorphism group $H(P_\infty)$ of the Hermitian function field $Herm|F_{q^2}$ acting on it as $\psi(y),\psi(z)$ has a $ordH(P_\infty)=q^3(q^2-1)$ greater than the order Suzuki group. The basic idea of MST cryptosystems is given in [9]. We have described the idea in details in our previous work in details [14-17].

## PROPOSED APPROACH

So, within this proposal we have the following steps of key generation, encryption and decryption.

*As a input we have* a large group $H(P_\infty)$. This group is based on the automorphism $\psi(y),\psi(z)$. The construction of group elements $H(P_\infty)$ is determined by solving the equation $\gamma^q+\gamma=\beta^{q+1}$ with respect to $\gamma$. The difficulty of finding $c$ is proportional to $q$. $H(P_\infty)$ of the $Herm|F_{q^2}$ can be represented by

$$H(P_\infty)=\left\{\left[\alpha,\beta,\frac{\beta^{q+1}}{2}+\gamma\right]\middle|\alpha\in F_{q^2}^*,\beta\in F_{q^2},\ \gamma^q+\gamma=0\right\}.$$

And it's true for the odd characteristic. If $\lambda$ is a generating element of the field, then the equation $\gamma^q+\gamma=0$ has solutions $\gamma_i=\lambda^{(q+1)/2+k(q+1)}$, $k=\overline{0,q-1}$. Computation vectors using logarithmic signatures matrices and random covers are now easily transcoded into the coordinates $\beta,\gamma$ of the $H(P_\infty)$ subgroup.

The group operation is defined as

$$S(\alpha_1,\beta_1,\gamma_1)\cdot S(\alpha_2,\beta_2,\gamma_2)=S(\alpha_1\alpha_2,\alpha_2\beta_1+\beta_2,\alpha_2^{q+1}\gamma_1+\alpha_2\beta_2^q\beta_1+\gamma_2).$$

The inverse of

$$S(\alpha,\beta,\gamma)^{-1} = S(\alpha^{-1},-\alpha^{-1}\beta,-\alpha^{-(g+1)}\gamma+\alpha^{-(g+1)}\beta^{q+1}).$$

Calculation of the inverse element $S(\alpha,\beta,\gamma)^{-1}$ in this representation expands the scope for $\gamma_1 = \frac{\beta_1^{q+1}}{2}+\gamma'_1$ and $\gamma_2 = \frac{\beta_2^{q+1}}{2}+\gamma'_2$. This is a key idea in constructing a logarithmic signature based on the $H(P_\infty)$ of the $Herm|F_{q^2}$. In another case, if $\gamma'_1$ and $\gamma'_2$ are solutions of equation $\gamma^q+\gamma=0$, the inverse element is strictly defined through expression $S(\alpha,\beta,\gamma)^{-1} = S(\alpha^{-1},-\alpha^{-1}\beta,\alpha^{-(q+1)}\gamma^q)$.

As an output we have a $[w,\gamma,f]$ as a open key with corresponding $[v,(\tau_0,...,\tau_s)]$ is used as a secret key.

### KEY GENERATION

Step 1. Choose a 1st tame logarithmic signature $v_{(1)} = [V_{1(1)},...,V_{s(1)}] = (v_{kn})_{(1)} = S(1,v_{kn(1)},v_{kn(1)}^{q+1}/2)$ of type $(r_{1(1)},...,r_{s(1)})$, $k=\overline{1,s(1)}$, $n=\overline{1,r_{i(1)}}$, $v_{kn(1)} \in F_{q^2}$.

Step 2. Choose a 2nd tame logarithmic signature $v_{(2)} = [V_{1(1)},...,V_{s(2)}] = (v_{kn})_{(2)} = S(1,0,v_{kn(2)})$ of type $(r_{1(2)},...,r_{s(2)})$, $k=\overline{1,s(2)}$, $n=\overline{1,r_{i(2)}}$, $v_{kn(2)} \in F_q \subset F_{q^2}$.

Step 3. Select a 1st random cover $w_{(1)} = [W_{1(1)},...,W_{s(1)}] = (w_{kn})_{(1)} = S(w_{kn(1)_1},w_{kn(1)_2},(w_{kn(1)_2})^{q+1}/2)$ of the same type as $v_{(1)}$, where $w_{kn} \in H(P_\infty)$, $w_{kn(1)_1},w_{kn(1)_2} \in F_{q^2} \setminus \{0\}$.

Step 4. Select a 2nd random cover

$$w_{(2)} = [W_{1(2)},...,W_{s(2)}] = (w_{kn})_{(2)} = S(w_{kn(2)_1},w_{kn(2)_2},(w_{kn(2)_2})^{q+1}/2+w_{kn(2)_3})$$

of the same as $v_{(2)}$, where $w_{kn(2)_2},w_{kn(2)_3} \in F_q \setminus \{0\} \subset F_{q^2}$.

Step 5. Choose $\tau_{0(l)},\tau_{1(l)},...,\tau_{s(l)} \in H(P_\infty) \setminus Z$, $\tau_{i(l)} = S(\tau_{i(l)_1},\tau_{i(l)_2},(\tau_{i(l)_2})^{q+1}/2)$, $t_{i(l)_k} \in F^\times$, $i=\overline{0,s(l)}$, $l=\overline{1,2}$. Let's $\tau_{s(1)} = \tau_{0(2)}$.

Step 6. Construct a homomorphism $f_1$ defined by

$$f_1(S(w_1,w_2,w_2^{q+1}/2)) = S(1,w_2,w_2^{q+1}/2)$$

Step 7. Calculating $g_{(1)} = [g_{1(1)},...,g_{s(1)}] = (g_{kn})_{(1)} = \tau_{(k-1)(1)}^{-1} f_1((w_{kn})_{(1)})(v_{kn})_{(1)}\tau_{k(1)}$, $k=\overline{1,s(1)}$, $n=\overline{1,r_{i(1)}}$,

Where $f_1\left((w_{kn})_{(1)}\right)(v_{kn})_{(1)} = S\left(1, w_{kn(1)_2} + v_{kn(1)}, w_{kn(1)_2}^{q+1}/2 + w_{kn(1)_2} v_{kn(1)}^q + v_{kn(1)}^{q+1}/2\right)$.

Step 8. Define a homomorphism $f_2\left(S(w_1, w_2, w_2^{q+1}/2)\right) = S(1, 0, w_2)$.

Step 9. Calculating $g_{(2)} = [g_{1(2)}, ..., g_{s(2)}] = (g_{kn})_{(2)} = \tau_{(k-1)(2)}^{-1} f_2\left((w_{kn})_{(2)}\right)(v_{kn})_{(2)} \tau_{k(2)}$, $k = \overline{1, s(2)}$, $n = \overline{1, r_{i(2)}}$, where $f_2\left((w_{kn})_{(2)}\right)v(b_{kn})_{(2)} = S(1, 0, w_{kn(2)_2} + v_{kn(2)})$.

An open key equals to $[f_1, f_2, (w_l, g_l)]$, and a secret key equals to $[v_{(l)}, (\tau_{0(l)}, ..., \tau_{s(l)})]$, $l = \overline{1, 2}$.

So, as an input for the encryption we have a text $x \in H(P_\infty)$, $x = S(x_1, x_2, x_3)$ and the public key $[f_1, f_2, (w_l, g_l)]$, $l = \overline{1, 2}$.

## ENCRYPTION

$Q = (Q_1, Q_2)$, $Q_1 \in Z_{|F_{q^2}|}$, $Q_2 \in Z_{|\mathbb{Z}|}$ should be chosen randomly. Then, we calculate the following

$$y_1 = w'(Q) \cdot x = w_1'(Q_1) \cdot w_2'(Q_2) \cdot x,$$

$$y_2 = g'(Q) = g_1'(Q_1) \cdot g_2'(Q_2) = S\left(*, w_{(1)_1}(Q_1) + v_{(1)}(Q_1) + *, w_{(2)_2}(Q_2) + v_{(2)}(Q_2) + *\right).$$

Cross-calculations of $\tau_{0(l)}, ..., \tau_{s(l)}$ is used to defined $(*)$ components. And its used for third coordinate to be added in the product of $w_{(1)_1}(Q_1) + v_{(1)}(Q_1)$.

Calculating

$$y_3 = f_1(w_1'(Q_1)) = S(1, w_{(1)_2}(Q_1), *),$$

$$y_4 = f_2(w_2'(Q_2)) = S(1, 0, w_{(2)_2}(Q_2)).$$

Output $(y_1, y_2, y_3, y_4)$ of the message $x$.

So, in the next chapter we going to evaluate the main idea of our proposal.

# COMPUTATIONAL VALIDATION OF THE PROPOSED APPROACH

Lets check the consistency of the proposed approach.

Fix the finite field $F_{q^2}$, $q^2 = 3^6$, $g(z) = z^6 + 2z + 2$ and

$$H(P_\infty) = \left\{ \left[ \alpha, \beta, \frac{\beta^{q+1}}{2} + \gamma \right] \middle| \alpha \in F_{q^2}^*, \beta \in F_{q^2}, \gamma^q + \gamma = 0 \right\}.$$

Product of two matrices is used for the group operation

$$S(\alpha_1, \beta_1, \gamma_1) \cdot S(\alpha_2, \beta_2, \gamma_2) = S(\alpha_1\alpha_2, \alpha_2\beta_1 + \beta_2, \alpha_2^{q+1}\gamma_1 + \alpha_2\beta_2^q\beta_1 + \gamma_2).$$

$$S(a_1, b_1, c_1) \cdot S(a_2, b_2, c_2) = S(a_1a_2, a_2b_1 + b_2, a_2^{q+1}c_1 + a_2b_2^q b_1 + c_2),$$

where $\gamma_1 = \frac{\beta_1^{q+1}}{2} + \gamma'_1$, $\gamma_2 = \frac{\beta_2^{q+1}}{2} + \gamma'_2$.

The inverse element is determined as

$$S(\alpha, \beta, \gamma)^{-1} = S(\alpha^{-1}, -\alpha^{-1}\beta, \alpha^{-(q+1)}\gamma^q).$$

$S(1,0,0)$ is a triple and it is an identity.

Step 1. Let`s construct a tame logarithmic signature $v_{(1)} = [V_{1(1)},...,V_{s(1)}] = (v_{kn})_{(1)} = S(1, v_{i_{kn}(1)}, v_{kn(1)}^{q+1}/2)$ of type $(r_{1(1)},...,r_{s(1)})$, $k = \overline{1, s(1)}$, $n = \overline{1, r_{i(1)}}$, $v_{kn(1)} \in F_{q^2}$ for coordinate $\beta$ and $v_{(2)} = (v_{kn})_{(2)} = S(1, 0, v_{i_{kn}(2)})$ of type $(r_{1(2)},...,r_{s(2)})$, $k = \overline{1, s(2)}$, $n = \overline{1, r_{i(2)}}$, $v_{kn(2)} \in F_q \subset F_{q^2}$ for coordinate $\gamma$.

The logarithmic signatures $v_1$ and $v_2$ in a group representations define $v_{kn(1)}$ and $v_{kn(2)}$ coordinates. Types $(r_{1(l)},...,r_{s(l)})$ and logarithmic signature s $v_1$ and $v_2$ are chosen independently. Let`s LSs $v_1$ and $v_2$, as an example, have a types $(r_{1(1)}, r_{2(1)}, r_{3(1)}) = (3^3, 3^2, 3)$, $(r_{1(2)}, r_{2(2)}) = (3^2, 3)$ Arrays $v_{kn(1)}$ consists of three subarrays and $v_{kn(2)}$ have of two subarrays with a $r_i$ as a rows quantity. Any arrays fragmentation can be selected with the condition $\prod_{i=1}^{s(1)} r_i = 3^6$ for $v_{kn(1)}$ and $\prod_{i=1}^{s} r_i = 3^3$ for $v_{kn(2)}$, respectively. Each row $v_{kn}$ it`s an $F_{q^2}$ field element. We construct arrays of logarithmic signatures with the method shown in [18].

To fulfill and increase the security requirements of arrays $v_l$ we can use different cryptographic transformations. As an obvious case, we can simply add noise vectors, permute strings in subarrays $V_i$, merge of arrays $V_i$ or use matrix transformations. It helps to build two different logarithmic signatures $v_1 = [V_{1(1)}, V_{2(1)}, V_{3(1)}]$ and $v_2 = [V_{1(2)}, V_{2(2)}]$. The arrays of logarithmic signatures $v_1 = S(1, v_{kn(1)}, v_{kn(1)}^{q+1}/2)$ and $v_2 = S(1, 0, v_{kn(2)})$ in the group representation, define the coordinates $v_{kn(1)}$ and $v_{kn(2)}$, respectively. In the string and the field representation $v_1$ and $v_2$ has the following form (See Table 1).

Table 1- Field representation of $v_1$ and $v_2$

| $v_1=$ | $v_{kn(1)}$ | $S$ | $v_1=$ | $v_{kn(1)}$ | $S$ |
|---|---|---|---|---|---|
| $V_{1(1)}$ | **000**000 | 1,0,0 | $V_{2(1)}$ | 211**000** | $1,\alpha^{512},\alpha^{140}$ |
| | **1**00000 | $1,\alpha^0,\alpha^{364}$ | | 202**1**00 | $1,\alpha^{518},\alpha^{308}$ |
| | **2**00000 | $1,\alpha^{364},\alpha^{364}$ | | 120**2**00 | $1,\alpha^{330},\alpha^{140}$ |
| | 0**1**0000 | $1,\alpha^1,\alpha^{392}$ | | 120**0**10 | $1,\alpha^{179},\alpha^{280}$ |
| | **1****1**0000 | $1,\alpha^6,\alpha^{532}$ | | 201**1**10 | $1,\alpha^{700},\alpha^{308}$ |
| | **2****1**0000 | $1,\alpha^{237},\alpha^{448}$ | | 111**2**10 | $1,\alpha^{203},\alpha^{224}$ |
| | 0**2**0000 | $1,\alpha^{365},\alpha^{392}$ | | 011**0**20 | $1,\alpha^{715},\alpha^0$ |
| | **1****2**0000 | $1,\alpha^{601},\alpha^{448}$ | | 202**1**20 | $1,\alpha^{414},\alpha^{308}$ |
| | **2****2**0000 | $1,\alpha^{370},\alpha^{532}$ | | 202**2**20 | $1,\alpha^{280},\alpha^{196}$ |
| | 00**1**000 | $1,\alpha^2,\alpha^{420}$ | $V_{3(1)}$ | 22021**0** | $1,\alpha^{39},\alpha^0$ |
| | **1**0**1**000 | $1,\alpha^{79},\alpha^{392}$ | | 02100**1** | $1,\alpha^{172},\alpha^{84}$ |
| | **2**0**1**000 | $1,\alpha^{243},\alpha^{616}$ | | 01102**2** | $1,\alpha^{354},\alpha^{84}$ |
| | 0**1****1**000 | $1,\alpha^7,\alpha^{560}$ | $v_2=$ | $v_{kn(2)}$ | $S(1,0,v_{kn(2)})$ |
| | **1****1****1**000 | $1,\alpha^{474},\alpha^{532}$ | | | |
| | **2****1****1**000 | $1,\alpha^{512},\alpha^{140}$ | $V_{1(2)}$ | **000**000 | 1,0,0 |
| | 0**2****1**000 | $1,\alpha^{238},\alpha^{476}$ | | **1**00000 | $1,0,\alpha^0$ |
| | **1****2****1**000 | $1,\alpha^{12},\alpha^{700}$ | | **2**00000 | $1,0,\alpha^{364}$ |
| | **2****2****1**000 | $1,\alpha^{688},\alpha^{700}$ | | 0**1**0000 | $1,0,\alpha^1$ |
| | 00**2**000 | $1,\alpha^{366},\alpha^{420}$ | | **1****1**0000 | $1,0,\alpha^6$ |
| | **1**0**2**000 | $1,\alpha^{607},\alpha^{616}$ | | **2****1**0000 | $1,0,\alpha^{237}$ |
| | **2**0**2**000 | $1,\alpha^{443},\alpha^{392}$ | | 0**2**0000 | $1,0,\alpha^{365}$ |
| | 0**1****2**000 | $1,\alpha^{602},\alpha^{476}$ | | **1****2**0000 | $1,0,\alpha^{601}$ |
| | **1****1****2**000 | $1,\alpha^{324},\alpha^{700}$ | | **2****2**0000 | $1,0,\alpha^{370}$ |
| | **2****1****2**000 | $1,\alpha^{376},\alpha^{700}$ | $V_{2(2)}$ | **1****2**0000 | $1,0,\alpha^{601}$ |
| | 0**2****2**000 | $1,\alpha^{371},\alpha^{560}$ | | **1****1****1**000 | $1,0,\alpha^{474}$ |
| | **1****2****2**000 | $1,\alpha^{148},\alpha^{140}$ | | 0**2****2**000 | $1,0,\alpha^{371}$ |
| | **2****2****2**000 | $1,\alpha^{110},\alpha^{532}$ | | | |

Step 2. Construct random covers $w_l$, for the same type as $v_1$ and $v_2$

$$w_{(1)} = [W_{1(1)},...,W_{s(1)}] = (w_{kn})_{(1)} = S\left(w_{kn(1)_1}, w_{kn(1)_2}, (w_{kn(1)_2})^{q+1}/2\right),$$

$$w_{(2)} = [W_{1(2)},...,W_{s(2)}] = (w_{kn})_{(2)} = S\left(w_{kn(2)_1}, w_{kn(2)_2}, (w_{kn(2)_2})^{q+1}/2 + w_{kn(2)_3}\right),$$

where $w_{kn(1)_1}, w_{kn(1)_2} \in F_{q^2}\setminus\{0\}$, $w_{kn(2)_3} \in F_q\setminus\{0\} \subset F_{q^2}$, $k = \overline{1,s_{(l)}}$, $n = \overline{1,r_{k(l)}}$, $l = \overline{1,2}$.

These covers $w_l$ to be defined by three arrays $\left(w_{kn(l)_1}, w_{kn(l)_2}, w_{kn(l)_3}\right)$ with non-zero entries.

Step 3. Let's generate random covers $w_1 = [W_{1(1)}, W_{2(1)}, W_{3(1)}]$, $w_2 = [W_{1(2)}, W_{2(2)}]$. In the field representation $w_1 = S\left(w_{kn(1)_1}, w_{kn(1)_2}, w_{kn(1)_3}\right)$ and $w_2 = S\left(w_{kn(2)_1}, w_{kn(2)_2}, w_{kn(2)_3}\right)$ has the following form (See Table 2). $\tau_{0(l)}, \tau_{1(l)},..., \tau_{s(l)} \in H(P_\infty)\setminus Z$, $\tau_{i(l)} = S\left(\tau_{i(l)_1}, \tau_{i(l)_2}, (\tau_{i(l)_2})^{q+1}/2\right)$, $\tau_{i(l)_k} \in F^\times$, $i = \overline{0,s(l)}$, $l = \overline{1,2}$ will be chosen randomly. Let's $\tau_{s(1)} = \tau_{0(2)}$.

Table 2- Field representation of $w_1$ and $w_2$

| $w_1$ | | | $w_2$ |
|---|---|---|---|
| $W_{1(1)}$ | $W_{1(1)}$ | $W_{2(1)}$ | $W_{1(2)}$ |
| $\alpha^{261},\alpha^{135},\alpha^{504}$ | $\alpha^{24},\alpha^{75},\alpha^{280}$ | $\alpha^{371},\alpha^{66},\alpha^{28}$ | $\alpha^{12},\alpha^{130},\alpha^{364}$ |
| $\alpha^{5},\alpha^{85},\alpha^{560}$ | $\alpha^{126},\alpha^{267},\alpha^{560}$ | $\alpha^{577},\alpha^{573},\alpha^{392}$ | $\alpha^{536},\alpha^{68},\alpha^{84}$ |
| $\alpha^{689},\alpha^{137},\alpha^{560}$ | $\alpha^{145},\alpha^{416},\alpha^{364}$ | $\alpha^{432},\alpha^{507},\alpha^{0}$ | $\alpha^{494},\alpha^{438},\alpha^{252}$ |
| $\alpha^{712},\alpha^{492},\alpha^{308}$ | $\alpha^{217},\alpha^{573},\alpha^{392}$ | $\alpha^{464},\alpha^{346},\alpha^{588}$ | $\alpha^{77},\alpha^{31},\alpha^{504}$ |
| $\alpha^{275},\alpha^{138},\alpha^{588}$ | $\alpha^{595},\alpha^{312},\alpha^{364}$ | $\alpha^{194},\alpha^{54},\alpha^{420}$ | $\alpha^{674},\alpha^{523},\alpha^{448}$ |
| $\alpha^{540},\alpha^{309},\alpha^{280}$ | $\alpha^{523},\alpha^{364},\alpha^{364}$ | $\alpha^{267},\alpha^{561},\alpha^{56}$ | $\alpha^{421},\alpha^{11},\alpha^{672}$ |
| $\alpha^{459},\alpha^{715},\alpha^{0}$ | $\alpha^{578},\alpha^{106},\alpha^{420}$ | $\alpha^{723},\alpha^{9},\alpha^{616}$ | $\alpha^{363},\alpha^{425},\alpha^{616}$ |
| $\alpha^{60},\alpha^{195},\alpha^{0}$ | $\alpha^{192},\alpha^{446},\alpha^{476}$ | $\alpha^{148},\alpha^{603},\alpha^{504}$ | $\alpha^{47},\alpha^{136},\alpha^{532}$ |
| $\alpha^{517},\alpha^{576},\alpha^{476}$ | $\alpha^{689},\alpha^{721},\alpha^{168}$ | $\alpha^{209},\alpha^{271},\alpha^{672}$ | $\alpha^{59},\alpha^{420},\alpha^{476}$ |
| $\alpha^{586},\alpha^{709},\alpha^{560}$ | $\alpha^{358},\alpha^{494},\alpha^{364}$ | | |
| $\alpha^{328},\alpha^{687},\alpha^{672}$ | $\alpha^{587},\alpha^{287},\alpha^{392}$ | $W_{3(1)}$ | $W_{2(2)}$ |
| $\alpha^{687},\alpha^{313},\alpha^{392}$ | $\alpha^{661},\alpha^{597},\alpha^{336}$ | $\alpha^{574},\alpha^{672},\alpha^{252}$ | $\alpha^{400},\alpha^{363},\alpha^{336}$ |
| $\alpha^{677},\alpha^{584},\alpha^{700}$ | $\alpha^{446},\alpha^{407},\alpha^{112}$ | $\alpha^{119},\alpha^{167},\alpha^{672}$ | $\alpha^{422},\alpha^{635},\alpha^{672}$ |
| $\alpha^{347},\alpha^{579},\alpha^{560}$ | | $\alpha^{466},\alpha^{95},\alpha^{112}$ | $\alpha^{505},\alpha^{512},\alpha^{140}$ |

Step 4. Let for the 1st logarithmic signature $\beta_1$ we have

| $\tau_{0(1)}=(\alpha^{267},\alpha^{675},\alpha^{336})$ | $\tau^{-1}_{0(1)}=(\alpha^{461},\alpha^{44},\alpha^{140})$ |
|---|---|
| $\tau_{1(1)}=(\alpha^{281},\alpha^{23},\alpha^{280})$ | $\tau^{-1}_{1(1)}=(\alpha^{447},\alpha^{106},\alpha^{420})$ |
| $\tau_{2(1)}=(\alpha^{52},\alpha^{490},\alpha^{252})$ | $\tau^{-1}_{2(1)}=(\alpha^{676},\alpha^{74},\alpha^{252})$ |
| $\tau_{3(1)}=(\alpha^{660},\alpha^{225},\alpha^{112})$ | $\tau^{-1}_{3(1)}=(\alpha^{68},\alpha^{657},\alpha^{560})$ |

Step 5. Let for the 2nd logarithmic signature $\beta_2$ we have

| $\tau_{0(2)}=(\alpha^{660},\alpha^{225},\alpha^{112})$ | $\tau^{-1}_{0(2)}=(\alpha^{68},\alpha^{657},\alpha^{560})$ |
|---|---|
| $\tau_{1(2)}=(\alpha^{79},\alpha^{475},\alpha^{560})$ | $\tau^{-1}_{1(2)}=(\alpha^{649},\alpha^{32},\alpha^{532})$ |
| $\tau_{2(2)}=(\alpha^{115},\alpha^{415},\alpha^{336})$ | $\tau^{-1}_{2(2)}=(\alpha^{613},\alpha^{664},\alpha^{28})$ |

The arrays $g_1$ and $g_2$ to be calculated within the next step.

Step 6. So, we obtain $g_{(1)} = [g_{1(1)},...,g_{s(1)}] = (g_{kn})_{(1)} = \tau^{-1}_{(k-1)(1)} f_1((w_{kn})_{(1)})(v_{kn})_{(1)} \tau_{k(1)}$,
$g_{(2)} = [g_{1(2)},...,g_{s(2)}] = (g_{kn})_{(2)} = \tau^{-1}_{(k-1)(2)} f_2((w_{kn})_{(2)})(v_{kn})_{(2)} \tau_{k(2)}$. by the condition of the example.

Step 7. Construct a homomorphism $f_1$, $f_2$ defined by
$f_1(S(w_1,w_2,w_2^{q+1}/2)) = S(1,w_2,w_2^{q+1}/2)$, $f_2(S(w_1,w_2,w_2^{q+1}/2)) = S(1,0,w_2)$.

In the field representation $g_1 = S(g_{kn(1)_1}, g_{kn(1)_2}, g_{kn(1)_3})$ and $g_2 = S(g_{kn(2)_1}, g_{kn(2)_2}, g_{kn(2)_3})$ has the following form (See Table 3)

Table 3- Field representation of $g_1$ and $g_2$

| $g_1$ | | | $g_2$ |
|---|---|---|---|
| $g_{1(1)}$ | $g_{1(1)}$ | $g_{2(1)}$ | $g_{1(2)}$ |
| $\alpha^{14},\alpha^{549},\alpha^{692}$ | $\alpha^{14},\alpha^{514},\alpha^{606}$ | $\alpha^{499},\alpha^{219},\alpha^{35}$ | $\alpha^{147},\alpha^{149},\alpha^{208}$ |
| $\alpha^{14},\alpha^{150},\alpha^{232}$ | $\alpha^{14},\alpha^{173},\alpha^{101}$ | $\alpha^{499},\alpha^{492},\alpha^{308}$ | $\alpha^{147},\alpha^{149},\alpha^{455}$ |
| $\alpha^{14},\alpha^{420},\alpha^{179}$ | $\alpha^{14},\alpha^{158},\alpha^{194}$ | $\alpha^{499},\alpha^{599},\alpha^{500}$ | $\alpha^{147},\alpha^{149},\alpha^{477}$ |
| $\alpha^{14},\alpha^{252},\alpha^{176}$ | $\alpha^{14},\alpha^{468},\alpha^{46}$ | $\alpha^{499},\alpha^{488},\alpha^{520}$ | $\alpha^{147},\alpha^{149},\alpha^{262}$ |
| $\alpha^{14},\alpha^{215},\alpha^{467}$ | $\alpha^{14},\alpha^{421},\alpha^{226}$ | $\alpha^{499},\alpha^{569},\alpha^{187}$ | $\alpha^{147},\alpha^{149},\alpha^{371}$ |
| $\alpha^{14},\alpha^{613},\alpha^{678}$ | $\alpha^{14},\alpha^{321},\alpha^{652}$ | $\alpha^{499},\alpha^{561},\alpha^{678}$ | $\alpha^{147},\alpha^{149},\alpha^{55}$ |
| $\alpha^{14},\alpha^{145},\alpha^{678}$ | $\alpha^{14},\alpha^{342},\alpha^{72}$ | $\alpha^{499},\alpha^{48},\alpha^{653}$ | $\alpha^{147},\alpha^{149},\alpha^{31}$ |
| $\alpha^{14},\alpha^{646},\alpha^{241}$ | $\alpha^{14},\alpha^{393},\alpha^{349}$ | $\alpha^{499},\alpha^{575},\alpha^{452}$ | $\alpha^{147},\alpha^{149},\alpha^{183}$ |
| $\alpha^{14},\alpha^{620},\alpha^{264}$ | $\alpha^{14},\alpha^{43},\alpha^{522}$ | $\alpha^{499},\alpha^{522},\alpha^{659}$ | $\alpha^{147},\alpha^{149},\alpha^{328}$ |
| $\alpha^{14},\alpha^{37},\alpha^{676}$ | $\alpha^{14},\alpha^{221},\alpha^{244}$ | | |
| $\alpha^{14},\alpha^{573},\alpha^{636}$ | $\alpha^{14},\alpha^{510},\alpha^{534}$ | $g_{3(1)}$ | $g_{2(2)}$ |
| $\alpha^{14},\alpha^{638},\alpha^{301}$ | $\alpha^{14},\alpha^{498},\alpha^{250}$ | $\alpha^{608},\alpha^{714},\alpha^{474}$ | $\alpha^{36},\alpha^{697},\alpha^{450}$ |
| $\alpha^{14},\alpha^{414},\alpha^{202}$ | $\alpha^{14},\alpha^{671},\alpha^{634}$ | $\alpha^{608},\alpha^{24},\alpha^{632}$ | $\alpha^{36},\alpha^{697},\alpha^{24}$ |
| $\alpha^{14},\alpha^{614},\alpha^{515}$ | | $\alpha^{608},\alpha^{149},\alpha^{180}$ | $\alpha^{36},\alpha^{697},\alpha^{380}$ |

For example, let $Q_1 = 379$. We obtain the following basis factorization for a given type $(r_{1(1)}, r_{2(1)}, r_{3(1)}) = (3^3, 3^2, 3)$ in the form of $Q_1 = (Q_{1(1)}, Q_{2(1)}, Q_{3(1)}) = (1,5,1)$, where $Q_1 + Q_2 3^3 + Q_3 3^5 = 379$.

Step 8. Calculating $g_1$
$$g_1(379) = g_{1(1)}(1) g_{2(1)}(5) g_{3(1)}(1) = S(\alpha^{14}, \alpha^{150}, \alpha^{232}) S(\alpha^{499}, \alpha^{561}, \alpha^{678}) S(\alpha^{608}, \alpha^{24}, \alpha^{632}) = S(\alpha^{393}, \alpha^{91}, \alpha^0).$$

Let $R_2 = 17$. We obtain the $Q_2 = (Q_{1(2)}, Q_{2(2)}) = (8,1) = 17$ for a given type $(r_{1(2)}, r_{2(2)}) = (3^2, 3)$.

Step 9. Calculating $g_2$
$$g_2(17) = g_{1(2)}(8) g_{2(2)}(1) = S(\alpha^{147}, \alpha^{149}, \alpha^{328}) S(\alpha^{36}, \alpha^{697}, \alpha^{24}) = S(\alpha^{183}, \alpha^{192}, \alpha^{433}).$$

In this section we consider step-by-step encryption algorithm. We have a message $x \in N(P_\infty)$, $x = S(x_1, x_2, x_3)$, $x_1 \in F_{q^2} \setminus \{0\}$, $x_2, x_3 \in F_{q^2}$ and the public key $[f_1, f_2, (w_l, g_l)]$, $l = \overline{1,2}$ for the input.

Step 10. Let $x = (\alpha^1, \alpha^2, \alpha^3) = S(\alpha^1, \alpha^2, \alpha^3)$.

$Q = (Q_1, Q_2) = (379, 17)$, $Q_1 \in \mathbb{Z}_{|F_{q^2}|}$, $Q_2 \in \mathbb{Z}_{|F_q|}$ to be chosen randomly.

Step 11. Calculating
$$y_1 = w'(Q) \cdot x = w_1'(Q_1) \cdot w_2'(Q_2) \cdot x = S(\alpha^{145}, \alpha^{602}, \alpha^{329}),$$
$$y_2 = g'(Q) = g_1'(Q_1) \cdot g_2'(Q_2) = S(\alpha^{576}, \alpha^{370}, \alpha^{226}),$$
$$y_3 = f_1(w_1'(Q_1)) = S(\alpha^0, \alpha^{394}, \alpha^{383}),$$
$$y_4 = f_2(w_2'(Q_2)) = S(\alpha^0, 0, \alpha^{692}).$$

Output a ciphertext $(y_1, y_2, y_3, y_4)$ of the message $x$.

Next, we consider step-by-step decryption algorithm. We have ciphertext $(y_1, y_2, y_3, y_4)$ and private key $[v_l, (\tau_{0(l)}, ..., \tau_{s(l)})]$ as an inputs.

Step 12. The random numbers $Q = (Q_1, Q_2)$ will be restored with the next steps to decrypt a message $x$:
$$D^{(1)}(Q_1, Q_2) = \tau_{0(1)} y_2 \tau_{s(2)}^{-1} = \tau_{0(1)} S(\alpha^{576}, \alpha^{370}, \alpha^{226}) \tau_{s(2)}^{-1} = S(\alpha^0, \alpha^{273}, \alpha^{139}),$$
$$D^*(Q) = y_3^{-1} D^{(1)}(Q_1, Q_2) = S(\alpha^0, \alpha^{30}, \alpha^{149}) S(\alpha^0, \alpha^{273}, \alpha^{139}) = S(\alpha^0, \alpha^{32}, \alpha^{408}).$$

We get $v_1(Q_1) = \alpha^{32} = (202211)$.

Step 13. Recovery of $Q_1$ was done earlier $Q_1 = (Q_{1(1)}, Q_{2(1)}, Q_{3(1)}) = (1,5,1)$.

```
      202|21|1                    Q1=(*,*,1)
      021|00|1                    row 1 from V3(1)
      202|21|1−021|00|1=211|21|0  Q1=(*,5,1)
      111|21|0                    row 5 from V2(1)
      211|21|0−111|21|0=100|00|0  Q1=(1,5,1)
```

Step 14. The components of the arrays $w_1'(Q_1)$ and $w_1'(Q_1)$ will be removed from ciphertext $(y_1, y_2)$ for further calculations:
$$y_2^{(1)} = \gamma_1'(Q_1)^{-1} y_2 = S(\alpha^{393}, \alpha^{91}, \alpha^0)^{-1} S(\alpha^{576}, \alpha^{370}, \alpha^{226}) = S(\alpha^{183}, \alpha^{192}, \alpha^{433}).$$

Step 15. Repeat the calculations
$$D^{(2)}(Q_2) = \tau_{0(2)} y_2^{(1)} \tau_{s(2)}^{-1} = \tau_{0(2)} S(\alpha^{183}, \alpha^{192}, \alpha^{433}) \tau_{s(l/2)}^{-1} = S(\alpha^0, 0, \alpha^{589}),$$

$$D^*(Q) = D^{(2)}(Q_2) y_4^{-1} = S(\alpha^0, 0, \alpha^{589}) S(\alpha^0, 0, \alpha^{692})^{-1} = S(\alpha^0, 0, \alpha^2).$$

Step 16. Restore $Q_2$ with $v_2(Q_2) = \alpha^2 = (001000)$.

Step 17. Perform inverse calculations $v_2(Q_2)^{-1}$. We found the bit groups in $v(Q)$ in accordance with a type $(r_{1(2)}, ..., r_{s(2)}) = (3^2, 3)$. Same computation to be used as in the example for $v_1(Q_1)^{-1}$. Then, we achieve

```
      00|1|000                                  Q₂=(*,1)
      11|1|000                                  row 1 from V₂₍₂₎
      00|1|000−11|1|000=22|0|000   Q₂=(8,1)
      Q₂ = (Q₁₍₂₎, Q₂₍₂₎) = (8,1) = 17
```

Step 18. Recover a message
$$x = w'(Q)^{-1} y_1 = [w_1'(Q_1) \cdot w_2'(Q_2)]^{-1} \cdot y_1 = [S(\alpha^{391}, \alpha^{39}, \alpha^{36}) S(\alpha^{481}, \alpha^{52}, \alpha^{637})]^{-1} S(\alpha^{145}, \alpha^{602}, \alpha^{329}) = S(\alpha^1, \alpha^2, \alpha^3)$$

*Output*: the message $x = (\alpha^1, \alpha^2, \alpha^3)$.

## SECURITY ANALYSIS

Let's consider and describe possible attacks.

First, the attack known as a brute force attack can be executed on ciphertext by selecting $Q = (Q_1, Q_2)$ with attempt to decipher the text $y_1 = w'(Q) \cdot x = w_1'(Q_1) \cdot w_2'(Q_2) \cdot x$. In this case, attack complexity is equal to $q^3$.

Second, we can also have a brute force attack on $Q = (Q_1, Q_2)$ by selecting such $Q = (Q_1, Q_2)$ to find $y_2 = g'(Q) = g_1'(Q_1) \cdot g_2'(Q_2)$. In this case, attack complexity is equal to $q^3$.

The third one is to choose $Q_1$ to match the value of $w_{(1)_2}(Q_1)$ in the vector $y_3 = f_1(w_1'(Q_1)) = S(1, w_{(1)_2}(Q_1), *)$. In this case, attack complexity is equal to $q^2$. Also, we can try to choose $Q_2$ to match the value of $w_{(2)_2}(Q_2)$ in the vector $y_4$. In this case, its less complex and equal to $q$. We consider using matrix transformation as a possible protection mechanism. So, link $Q_1$ and $Q_2$ it would help with this.

Next, we can apply brute force attack on the $(\tau_{0(l)}, ..., \tau_{s(l)})$ vectors. In this case, attack complexity is equal to $(q^2)^2$.

Also, there is an attack on the algorithm by itself. Extraction parameters of $y_3$, $y_4$ does not allow to calculate $w_1'(Q_1) \cdot w_2'(Q_2)$ in $y_1 = w_1'(Q_1) \cdot w_2'(Q_2) \cdot x$. If we simply try to find the parameters $Q_1, Q_2$, we need efforts equal to a brute force attack with complexity $q^2$. Since the $H(P_\infty)$ of $Herm|F_{q^2}$ is defined over a field $F_{q^2}$ which is large enough then this attack is just not feasible.

## CONCLUSION

Construction of a logarithmic signature $\beta$ with specific parameters is required for the cryptosystem based on $H(P_\infty)$ of the $Herm|F_{q^2}$. Here, the logarithmic signature $v = [V_1, ..., V_s] = (v_{kn}) = S(1, v_{kn(2)}, v_{kn(3)})$ Is a subgroup of the $H(P_\infty) = \{S(\alpha, \beta, \gamma) | \alpha, \beta \in F_{q^2}, \gamma^q + \gamma = \beta\}$. It's the same for a random cover $w = [W_1, ..., W_s] = (w_{kn}) = S(w_{kn(1)}, w_{kn(2)}, w_{kn(3)})$. It has the same type as $v$. In fact, the size of the array $v$ and $w$ is defined by the type $(r_1, ..., r_s)_2$ and $(r_1, ..., r_s)_3$ for $\beta, \gamma$ in the $H(P_\infty)$ subgroups. Thus, they both have to be converted to the elements of the groups. The solution to the problem is found for the case when the field has an odd characteristic. It also requires being on the extension of the automorphism group.

The $H(P_\infty)$ of the $Herm|F_{q^2}$ has a simple representation for the odd characteristic field. The vectors using logarithmic signature matrices and random covers are now easily transcoded. And it gives us coordinates of the $H(P_\infty)$ subgroup. This gives an advantage in the size of the message for the proposed MST3 cryptosystem design.